\documentclass[preprint2]{aastex}
\usepackage{graphics}
\usepackage{amssymb}
\pdfoutput=1
\usepackage{textcomp}
\usepackage{array}
\usepackage{multirow}
\usepackage{color}

\shorttitle{A two-phase scenario for bulge assembly in $\Lambda$CDM cosmologies}
\shortauthors{Obreja et al.}

\begin{document}

\title{A two-phase scenario for bulge assembly \\ 
    in $\Lambda$CDM cosmologies}

\author{A. Obreja, R. Dom\'{\i}nguez-Tenreiro, C. Brook}
\affil{Depto. de F\'{i}sica Te\'orica, Universidad Aut\'onoma de Madrid, 28049 Cantoblanco Madrid, Spain}
\email{aura.obreja@uam.es}

\author{F.~J. Mart\'{\i}nez-Serrano, M. Dom\'{e}nech-Moral, A. Serna}
\affil{Depto. de F\'{i}sica y Arquitectura de Computadores, Universidad Miguel Hern\'{a}ndez, 03202 Elche, Spain}

\author{M. Moll\'{a}}
\affil{Depto. de Investigaci\'{o}n B\'{a}sica, CIEMAT, 28040 Madrid, Spain}

\and 

\author{G. Stinson}
\affil{Max-Planck-Institut f\"ur Astronomie, K\"onigstuhl 17, 69117, Heidelberg, Germany}

\begin{abstract}
We analyze  and compare the bulges of a sample of $L_*$ spiral galaxies in hydrodynamical simulations in a cosmological context, using two different codes, {\tt P-DEVA} and {\tt GASOLINE}. 
The codes regulate star formation in very different ways, with {\tt P-DEVA} simulations inputing low star formation efficiency under the assumption that feedback occurs on subgrid scales, 
while the {\tt GASOLINE} simulations have feedback which drives large scale outflows. 
In all cases, the marked knee-shape in mass aggregation tracks, corresponding to the transition from an early phase of rapid mass assembly to a later slower one, 
separates the properties of two populations within the simulated bulges. 
The  bulges analyzed show an important early starburst resulting from the collapse-like fast phase of mass assembly, followed by a
 second phase with lower star formation, driven by a variety of processes such as disk instabilities and/or mergers.  
Classifying bulge stellar particles identified at $z=0$ into old and young according 
to these two phases, we found bulge stellar sub-populations with distinct 
 kinematics, shapes, stellar ages and metal contents. The young components are more oblate, generally smaller, more rotationally supported, with higher metallicity and less  
alpha-element enhanced than the old ones. 
These results  are consistent with the current observational status of  bulges,
and provide an explanation for some apparently paradoxical observations, such as bulge rejuvenation and metal-content gradients observed.
Our results suggest that bulges of $L_*$ galaxies will generically have two bulge populations which can be likened to classical and pseudo-bulges, 
with differences being in the relative proportions  of the two, which may vary due to galaxy mass and specific mass accretion and merger histories.
\end{abstract}


\section{Introduction}
\label{intro}

Understanding how bulges form and evolve is important  within the 
theories and models of galaxy formation and evolution.
Bulges are the component responsible for the central light in excess of the  exponential disk \citep{Freeman:1970}, 
and account for more than 25\% of the starlight  emitted  in the local Universe. 
There are three main classes of observed bulge: classical, pseudo-,
and boxy or peanut-shaped bulges, see \cite{Kormendy:2004} and \cite{Athanassoula:2005} for a 
detailed discussion on bulge classification. 
The distinction is important in that different classes cloud have formed and evolved in different ways.

The Milky Way Bulge is the only case in which individual stars are resolved, and thus provides unique information about bulge  properties. 
The Milky Way is considered to have a boxy bulge, yet increased evidence for an old, $\alpha$-enriched stellar population 
formed on a short time-scale has resulted in a  two-component model \citep[e.g.][]{Tsujimoto:2012} of the Bulge.
It had been shown  that two stellar populations  coexist in the Bulge, separated in age and metallicity
\citep{Mcwilliam:1994, Feltzing:2000, Barbuy:1999, vanLoon:2003, Groenewegen:2005, Zoccali:2006, Fulbright:2007, Zoccali:2008} and that the separation somewhat extends to kinematics 
\citep{Zhao:1994, Soto:2007}, even if  age determinations through color-magnitude diagrams showed that most bulge 
stars in the Galaxy are older than 10 Gyrs \citep{Ortolani:1995, Feltzing:2000, Zoccali:2006, Clarkson:2008}.

Recently \citet{Bensby:2011} reported a 2-13 Gyrs age span among microlensed turn-off Bulge stars. 
These latest data from high resolution spectrography, have put increased emphasis on the picture of a two-component Bulge.
Indeed, analyses of giant stars confirmed the presence of two distinct populations: 
a metal-poor enriched in [$\alpha$/Fe] one with kinematics consistent with an old spheroid, and a metal-rich one with roughly solar [$\alpha$/Fe] 
and with bar-like kinematics \citep{Babusiaux:2010, Hill:2011, Gonzalez:2011, dePropris:2011, Johnson:2011, Robin:2012, Soto:2012, Ness_conf:2012, Ness:2012}. 
Comparing metallicities and compositions at different galactic latitudes, it has been found that while the [$\alpha$/Fe] remains roughly constant and metal-poor stars
show a remarkable homogeneity in the Bulge \citep{Lecureur:2007, Johnson:2011, Gonzalez:2011}, the most metal-rich stars near the galactic plane disappear  
at higher latitudes \citep[see][among others]{Zoccali:2008,Lecureur:2007,Babusiaux:2010}. Further, the younger population is associated with the bar \citep[][and references therein]{Babusiaux:2010}. 
These differences could be an indication that younger, more metal-rich stars in the Bulge define a smaller region than metal-poor ones \citep[see also][]{Robin:2012}. 
Otherwise, \citet{Johnson:2012} explored the link between Bulge and thick disk formation and found diverging behaviors in the cases of [Na/Fe] and [La/Fe], 
which results in the Bulge resembling more the spheroid.

In the bulges of external galaxies, similar conclusions have been reached. Stellar population studies suggest that in many cases a secondary stellar population is superimposed on an older one 
\citep{Ellis:2001, Thomas:2006, Carollo:2007}, and that the two populations are  kinematically distinguishable, 
with the old population having spheroid-like kinematics, while this secondary population is more disk-like, see for example \citet{Prugniel:2001}, and the results from the SAURON 
project \citep{Peletier:2007, Erwin:2008}.  Bulge stellar masses are generally dominated by the old populations, with the young ones contributing less than 
a 25\% in most cases \citep{MacArthur:2009}, although \cite{Kormendy:2010} points out that significant numbers of local massive spiral galaxies appear to have dominant pseudo-bulges.

\citet{Jablonka:2007} conclude that most external bulges in their observed sample are more metal rich and have lower [$\alpha$/Fe] enhancement 
in their central regions than in their outer parts, a result consistent with \cite{Moorthy:2006},
and with Milky Way Bulge results. 
\citet{MacArthur:2009} find a wide range of gradients, both positive and negative, allowing for different bulge formation mechanisms. 
A complementary piece of information are age gradients, where  most authors find that the central regions of external bulges are younger than the outer ones, 
 \citep[e.g.][]{Moorthy:2006,Jablonka:2007,MacArthur:2009,SanchezBlazquez:2011}. Note, however, that  the last authors have also found negative age gradients in some cases.

Summarizing,  it would appear that bulge formation and evolution in a cosmological context has to account for a duality in stellar populations, 
and  for the coexistence of spheroid-like features with disk-like ones,  whilst some particularities may be related to minor merger events \citep[see ussion in][]{Combes:2009}.
Theoretical models focusing on the physical processes responsible for  bulge properties have a long history, a sample of which is presented below.
 Generally, the classical type is thought to originate from fast gas collapse at high redshift, or from gas clumps in a proto-disk that are driven to the central regions by dynamical friction. 
Whilst instabilities of the disk, perhaps associated with  bars, are believed  to result in  the formation of pseudo- and/or boxy bulges.

Metal enrichment in  bulges was first analytically studied through pure chemical evolution models. 
\citet{Matteucci:1990} first predicted an [$\alpha$/Fe] enhancement at bulges as a consequence of the different SNe I and II nucleosynthetic time-scales, 
observationally confirmed by  \citet{Mcwilliam:1994}. \citet{Molla:2000} used a multiphase multi-zone chemical evolution model and provided solutions 
to bulge chemical abundances and spectral indices. \citet{Ferreras:2003} found that very short infall time-scales are required for bulges. 
These models provided the basis of later developments to be implemented in hydrodynamical simulations, and, at the same time, gave us 
clear insights into some of the physical processes involved. 

On the other hand, purely collisionless numerical studies of bulge formation have focused on morphology and dynamics by analyzing disk evolution in pre-prepared simulations. 
Such studies showed that  spheroidal-like bulges cannot be formed through bar-buckling instabilities \citep{Debattista:2004} while disk-like and boxy/peanut bulges are different in their nature, 
although both classes are associated with the presence of a bar \citep{Athanassoula:2005}. 
On the other hand, \citet{ElicheMoral:2011} analyze the effect of minor mergers on the inner part of disk galaxies, finding this process to be efficient in forming 
rotationally supported stellar inner components, i.e. disks, rings or spiral patterns.
\citet{Hopkins:2010} use  a different approach, 
by constructing  semi empirical models, based on observationally motivated halo occupation numbers from the Millenium simulation, 
aiming to  quantify  the relative effect of galaxy mergers on bulge formation. 
They find major mergers to be the dominant mechanism for L$_*$ bulge and spheroid formation and assembly, while minor mergers contribute relatively more in lower mass systems.

The first numerical studies of bulge formation using dissipative collapse  by \citet{Samland:2003}  model the  formation of a large disk inside a spinning ($\lambda=0.05$), 
growing dark matter (DM) halo, with an added accretion history taken from the large-scale simulations of the GIF-VIRGO consortium \citep{Kauffmann:1999}. 
Chemical evolution is followed through two fiducial elements tracing the fraction of heavy elements produced by type Ia or type II SNe. 
The resulting bulge consists of at least two stellar sub-populations, an early collapse population and another one that formed later in the bar.
\citet{Nakasato:2003} presented results of the evolution of a spherical 3$\sigma$ top-hat over-dense region of 1.4 Mpc co-moving radius in rigid rotation 
evolved with a gravohydrodynamical code. Again the nucleosynthetic yields of type Ia or type II SNe are used, tracing the chemical enrichment in metallicity, O and Fe. 
Their results suggest that bulges consist of two chemically different components; one that has formed quickly through a sub-galactic merger in the proto-galaxy, 
and another one that formed gradually in the inner disk. \cite{Kobayashi:2011} further developed this latter chemical evolution implementation, and used it to simulate a Milky Way-like galaxy. 
Their kinematic and chemical results follow closely the observed properties of the Galaxy halo, bulge and thick disk. 
These simulations, however, do not take into account  the cosmological gas infall. 

Few studies on bulge formation have been made within  a fully cosmological context. 
\citet{Tissera:1998} and  \citet{Governato:2009} have studied  the effects of mergers on classical bulge stellar populations. 
\citet{Guedes:2011} have run the highest resolution up-to-now simulation of Milky Way like disk formation, 
and got realistic disk and bulge properties but their focus on  the details of bulge formation is relatively minor. 
\citet{Okamoto:2012} finds that pseudo bulges form, in his simulated Milky Way 
galaxies, by rapid gas supply at high-redshift, with their progenitors observable as high-redshift disks, 
and that  this  occurs  prior to formation of the final disk.
\citet{Brook:2012} obtained a lower mass late-type disk galaxy which has a  bulge that grows from $z=1$ mainly through purely secular processes.  
Other authors have put more emphasis on metal enrichment. For example, \cite{Rahimi:2010} run a fully cosmological simulation with the {\tt GCD+} code \citep{Kawata:2003},  
which incorporates chemical enrichment both by SNe Ia \citep{Iwamoto:1999,Kobayashi:2000} and SNe II \citep{Woosley:1995}, as well as mass loss from intermediate mass stars. 
The code does not include a mechanism to diffuse metals between gas particles, 
resulting in an artificially high spread in the metallicity distributions, but robust averages. 
Their results underline the importance of mergers in bulge formation and their possible kinematic implications, the dependence of metal content on age, 
and the  existence of accreted stars within the bulge.
  
Thus it would appear that a variety of conclusions are being drawn from different groups using different codes and different physical models for their  galaxy formation simulations. 
By analyzing in a unified manner (measuring the relevant properties with the same pipeline) simulated disk galaxies that are run with different codes and different physical 
prescriptions and merger histories, 
we hope in this study to shed light on what processes of bulge formation, and subsequent bulge properties, 
are common within such simulations. 
Specifically, we analyze the bulges of three of the more massive galaxies presented in \cite{Domenech:2012}, run with the {\tt P-DEVA} \citep{Serna:2003,Martinez:2008} code, 
and those of two {\tt GASOLINE} galaxies described in \cite{Brook:2012} and G.~Stinson et al. (2012, submitted), 
with the aim of deciphering the patterns of bulge formation by focusing in the properties of their stellar populations. 

\cite{Domenech:2012} analyze disk formation in a cosmological context by running zoom simulations with {\tt P-DEVA}. 
This code \citep{Martinez:2008} incorporates a statistical implementation of chemical enrichment based on \cite{Talbot:1973}, 
including chemical feedback both by SNe Ia and SNe II, as well as mass loss from intermediate mass stars, involving 11 elements. 
Radiative cooling takes into consideration the full element distribution at each point and time through a particular implementation of the Dimension Reduction regression \citep{Weisberg:2002}, 
while a SPH metal diffusion term mimics turbulent effects \citep{Monaghan:2005}. 
They have produced disk systems whose different components (thin and thick disk, halo, bulge) have properties nicely consistent with observations,
for example, they have g-band $B/T$ ratios between 0.13 and 0.36. 
In particular, bulge metallicity and [$\alpha$/Fe] distributions show bimodal patterns, that they interpret as resulting from fast and slow modes of star formation. 
The simulations run with {\tt P-DEVA} assume that supernova feedback works on sub-grid scales and 
results in a low star formation efficiency, which is thus used as an input parameter \citep[see  discussion in][]{Agertz:2011} that implicitly mimics the energetic feedback.

On the other hand, the simulations run with the  {\tt GASOLINE} code have  explicit feedback from massive stars  which drives large scale outflows. 
The two particular {\tt GASOLINE}  simulations were chosen for this study because (i) G-1578411 was shown to have a bulge which is mostly formed after the merger epoch. 
It is the late type simulated disk galaxy from \citet{Brook:2012}. A secular bulge grows between $z=1$ and $z=0$, driven at least partly by a bar. 
The final bulge to total light ratio is B/T=0.21.
(ii) G-1536 is the simulated L*  galaxy run by our MaGICC program that matches the widest range of observed galaxy properties 
(see G.~Stinson et al. (2012, submitted) and SG5LR in \citet{Brook:2012b}). 
It has  a B/T ratio of 0.35.

We use this suite of 5 inhomogeneous simulated disk galaxies to  analyze the mass-weighted three-dimensional shape, size and kinematics of the bulges, 
as well as their mass-weighted age, metallicity  and chemical composition. 
More specifically, we first show that the simulated bulges consist of roughly two stellar populations (old and young) 
whose properties are correlated with their shape, kinematics, metallicity and composition,  
in line with the observational results we have previously described. In view of this agreement with observations, we investigate the physical processes underlying bulge formation and 
particularly, their relation with the dynamics of the cosmic-web at high redshift, and of their host galaxies at low $z$s.

An  interesting aspect of this work is to clarify to what extent the 
formation mechanisms proposed for massive ellipticals also apply to the formation of bulges, 
as suggested by similarities in observed properties
\citep[e.g.][]{Franx:1993,Peletier:1999,Carollo:2001,Bureau:2002,Peletier:2008}.
In this regard, let us recall that the characteristics of mass assembly and star formation rate histories in ellipticals can  be interpreted in terms of dark halo dynamics and its consequences.
Indeed, analytical models as well as N-body simulations show that two different phases can be distinguished along the {\it halo} mass assembly process \citep{Wechsler:2002,Zhao:2003}:
i) first a violent fast phase with high mass aggregation rates, resulting from collapse-like and merger events, and ii) later on a slow phase with much lower mass aggregation rates.
Small box hydrodynamical simulations \citep{DT:2006} as well as larger box ones \citep{Oser:2010,DT:2011} confirmed this scenario as well as its implications on elliptical properties
at low $z$ \citep{Cook:2009}. This scenario nicely explains apparently paradoxical observational data of elliptical galaxies.

The paper is structured as follows.
In Section\,\ref{PDEVA} we describe the details of the simulations.
The criteria used to select bulge stars are discussed in Section\,\ref{select}.
In Section\,\ref{stepops}  we construct the star formation histories of the bulges and define the selection criteria for the old and young stellar components.
In the Section\,\ref{shape} we discuss their shapes and kinematics, and in Section\,\ref{chem} their chemical composition.
The origin of the bulge stars in the framework of host galaxy and cosmic-web dynamics is given in Section\,\ref{twophase}.
Finally, in Section\,\ref{summary} we draw our conclusions after a brief summary of our findings and discuss the
bulge formation scenario that emerges from our simulations.

\section{The simulations}
\label{PDEVA}
The main  properties and resolution of our five simulated galaxies are shown in Table~\ref{tab1}. 
All galaxies have previously  appeared in the literature, where more details can be garnered.  Here we outline the main features of the codes and simulations. 

\subsection{P-DEVA}

We use the OpenMP parallel version of the {\tt DEVA} code \citep{Serna:2003}, 
which includes the chemical feedback and cooling methods described in \citet{Martinez:2008}, and in which the conservation laws 
(e.g. momentum, energy, angular momentum and entropy) hold accurately \citep[][]{Serna:2003}. 
Full details of the recipes used are found in   \cite{Domenech:2012} where the simulations are first presented. 
The star formation recipe  follows a Kennincutt~\textendash~Schmidt-like law with a given density threshold, $\rho_*$, and star formation efficiency $c_{*}$.
In line with \citet{Agertz:2011},  we implement inefficient SF parameters (see Table \ref{tab1}), 
which implicitly account for the regulation of star formation by feedback energy processes by mimicking their effects, which are assumed to work on sub-grid scales.

The chemical evolution implementation \citep{Martinez:2008} accounts for the  dependence of radiative cooling on the detailed metal composition of the gas, 
by means of a fast algorithm based on a metallicity parameter, $\zeta(T)$. 
The code also tracks the full dependence of metal production on the detailed chemical composition of stellar particles \citep{Talbot:1973}, through
a $Q_{ij}$ formalism implementation of the stellar yields, for the first time in SPH. The delayed gas restitution from stars has been
 implemented through a probabilistic approach, that reduces statistical noise when compared with previous approaches, 
and therefore allows for a fair description of element enrichment
at a lower computational cost. Moreover, the metals are diffused in such a way as to mimic the turbulent mixing in the interstellar medium.

\begin{table*}
\begin{minipage}{6.3in}
\renewcommand{\thefootnote}{\thempfootnote}
\begin{center}
\begin{tabular}{|c|c|c|c|c|c|c|c|c|c|c|} \hline
Object & $\delta$M$_{bar}$ & h$_{soft}$ & $\rho_*$ & c$_*$ &IMF&E$_{SN}$& r$_{bulge}$ & M$_{*}$ & M$_{gas}$ & L$_{box}$\\
 & 10$^{5}$M$_\odot$ & h$^{-1}$kpc & cm$^{-3}$ & \% && 10$^{51}$ergs& kpc & 10$^{10}$M$_\odot$ & 10$^{8}$M$_\odot$ & Mpc\\ \hline\hline
G-1578411 & 0.2 & 0.15 & 9.4 & 1.7 &Kroupa93 &1.0 & 1.30 & 0.83 & 1.64 & 34\\
G-1536 & 1.9 & 0.15 & 9.4 & 3.3 &Chab03&1.0 & 2.10 & 0.96 & 5.92 & 68\\
HD-5004A & 3.94 & 0.2 & 6 & 1.0 &Salp55 &- &  1.00 & 1.50 & 3.67 & 10\\
HD-5004B & 3.94 & 0.2 & 10 & 0.8 &Salp55 &- & 1.55 & 1.60 & 5.33 & 10\\
HD-5103B & 3.78 & 0.2 & 12 & 0.8 &Salp55 &- & 1.73 & 1.47 & 4.61 & 10\\ \hline
\end{tabular}\\
\caption{
The initial mass of gas particles ($\delta$M$_{bar}$), minimum SPH smoothing length (h$_{soft}$), density threshold ($\rho_*$), star formation efficiency (c$_*$), 
initial mass function (IMF), SN feedback (E$_{SN}$),
bulge radius (r$_{bulge}$), stellar (M$_{*}$) and gas (M$_{gas}$) mass of the simulated bulges (at $z=0$), and periodic box length (L$_{box}$). 
The mass values correspond to the position selection. The {\tt P-DEVA} runs use a fixed mass for the baryonic particles, equal to $\delta$M$_{bar}$.}
\label{tab1}
\end{center}
\end{minipage}
\end{table*}

The simulations use the  cosmological "zoom-in" technique, with high-resolution gas and dark matter in the  region of the main object. The cosmological parameters of a $\Lambda$CDM model
were assumed ($\Omega_{\Lambda}=0.723$, $\Omega_m=0.277$, $\Omega_b=0.04$, and $h=0.7$), in a 10 Mpc per side periodic box.
Stellar masses are distributed according to the Salpeter initial mass function (IMF) \citep{Salpeter:1955}, with a mass range of [M$_l$,M$_u$]=[0.1,100]M$_\odot$. 

\subsection{GASOLINE}

The {\tt GASOLINE} galaxies are cosmological zoom simulations derived from the McMaster Unbiased Galaxy Simulations \citep[MUGS,][]{Stinson:2010}. 
In G-1578411, the initial conditions (ICs) are scaled down, so that rather than residing in a 68 h$^{-1}$ Mpc cube, 
it is inside a cube with 34 h$^{-1}$ Mpc sides, while G-1536 uses the same ICs as  in the MUGS runs.

When gas becomes cool ($T < 15000$ K) and dense ($n_{th} > 9.3$ cm$^{-3}$), 
it is converted to stars according to a Kennincutt~\textendash~Schmidt-like law with the star formation rate $\propto \rho^{1.5}$.
Effective star formation rates are determined by the combination and interplay of  $c_\star$ and feedback, 
$c_\star$ is ultimately the free parameter that sets the balance of the baryon cycle off cooling gas, star formation, and gas heating.
Stars feed energy back into surrounding gas. 
Supernova feedback is implemented using the blastwave formalism \citep{Stinson:2006} and deposits $10^{51}$ erg of energy 
into the surrounding medium at the end of the stellar lifetime of every star more massive than 8\,M$_\odot$. 
Energy feedback  from massive stars prior to their explosion as supernovae has also been included.
Without the ability to resolve the details, we  employ a  relatively crude thermal implementation of radiation feedback from massive stars, 
with the  aim being to mimic their most important effects on scales that we resolve, i.e. to regulate star formation, enhance inhomogeneity, 
and  to allow the expansion of  the SNe driven super-bubbles which drive outflows. 
To mimic the weak coupling of this energy to the surrounding gas \citep{Freyer:2006}, we inject pure thermal energy feedback, 
which is highly inefficient in these types of simulations  \citep{Katz:1992,Kay:2002}. We inject 10\% of the  available energy during this early stage of massive star evolution, 
but 90\% is rapidly radiated away, making an effective coupling of the order of 1\%.

The two simulations have different initial mass functions, with G-1536 having a factor of $\sim 2$ more massive stars \citep{Chabrier:2003} than G-1578411 \citep{Kroupa:1993}, 
meaning that G-1536 is less efficient at turning baryons into stars. 

Ejected mass and metals are distributed to the nearest neighbor gas particles using the smoothing kernel \citep{Stinson:2006}. 
Literature yields for SNII \citep{Woosley:1995} and SNIa \citep{Nomoto:1997} are used. 
Metal are diffused  by treating unresolved turbulent mixing as a shear-dependent diffusion term \citep{Shen:2010}, 
allowing proximate gas particles to mix their metals. Metal cooling is calculated based on the diffused metals.

\section{Bulge selection criteria}
\label{select}

\begin{figure*}[!ht]
\centering
\includegraphics[scale=1.0]{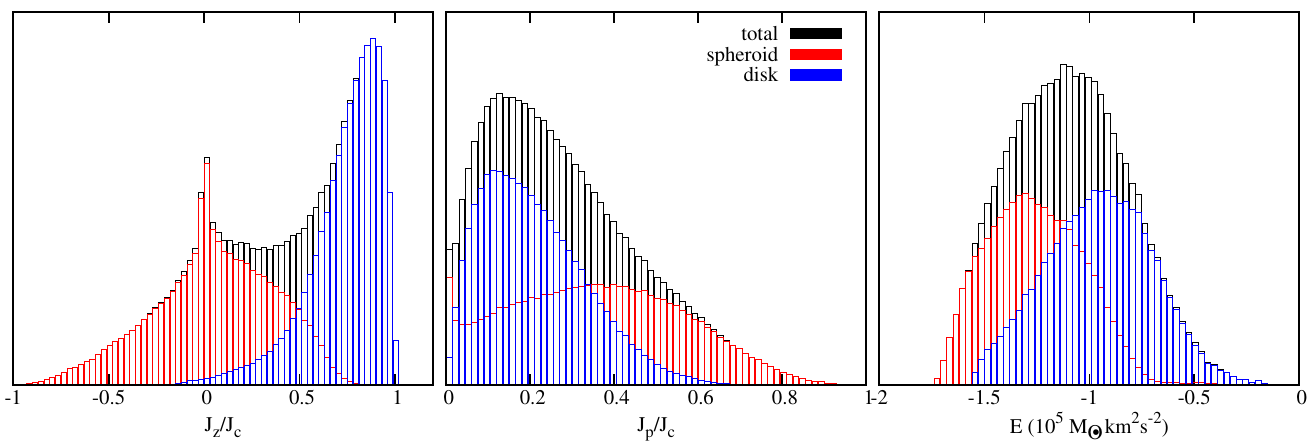}
\caption{The J$_z$/J$_c$ (left), J$_p$/J$_c$ (center) and $E$ (right) histograms for the galaxy G-1536. The spheroid and disk contributions, as derived 
with the clustering algorithm are given in red and blue, respectively. The histograms considering all the galaxy stellar particles are given in black.}
\label{cluster_hist}
\end{figure*}

\begin{figure}[!h]
\includegraphics[scale=1.0]{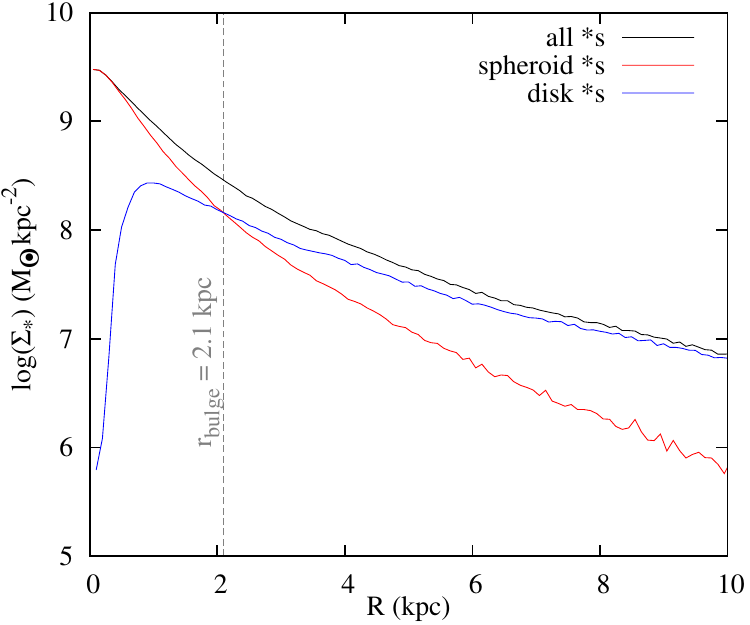}
\caption{The projected stellar mass density for the kinematic disk (blue curve) and spheroid (red curve), and all stellar particles (black curve) of galaxy G-1536. 
The grey dashed line gives the position of the radial cut used to define the bulge region.}
\label{mass_dens}
\end{figure}

The classical method of separating the contribution of the inner region from the disk rests on fitting 
the light radial profile of the spiral galaxy with two (or more) components (e.g. exponential disk + a S\'{e}rsic profile). 
However, this method does not provide any way in which individual stellar particles can be assigned to the inferred galaxy components. 
Therefore we make a kinematic separation of  bulge and  disk stars   using clustering algorithms.
For each star particle we computed  J$_p$, the angular momentum in the plane of the disk, J$_z$, the angular momentum perpendicular to the disk,  and J$_c$,  
the angular momentum  of a particle with the same binding energy (E), moving in a circular orbit at the same radius. 
Each stellar particle within the virial radius is then dynamically defined by three variables: J$_z$/J$_c$ , J$_p$/J$_c$, and  binding energy ($E$). 
These variables are normalized to [0,1] in order to give equal weight to all variables, and then fed to the clustering algorithm,
which requires as prior only the number of clusters (we chose {\it n=2} for bulge + disk). 
The normalization assumes a linear mapping between [X$_{min}$,X$_{max}$] and [0,1] (with J$_z$/J$_c$, J$_p$/J$_c$ and $E$ as X), 
which is appropriate if the histogram of X does not have extremely extended tails.
We considered the distance in the phase space to be the Euclidean one. The algorithm, starting from an initial random partitioning into {\it n} clusters, 
iteratively searches for the partitioning which would minimize the intracluster distance. 
In Figure~\ref{cluster_hist} we show the output of the clustering 
algorithm in the form of the histograms of J$_z$/J$_c$, J$_p$/J$_c$ and $E$ for disk, spheroid and galaxy stars in the case of G-1536. As expected, the spheroid stars 
peak at $J_z/J_c = 0$ in accordance with a system sustained by velocity dispersion, while the disk ones peak closer to $J_z/J_c = 1$. In the J$_p$/J$_c$ histrograms 
the disk shows a peak at $\sim$ 0.1, characteristic of a system whose particles move little perpendicular to the disk plane, although some of them can have an important 
vertical motion which shows up as the tail extending to $J_p/J_c \sim 0.6$.

Once the dynamical decomposition has been done, the projected stellar mass density profiles of the disk (thin plus thick) and spheroid naturally provide us with the radial cut necessary 
to delimit the bulge, in the form of the radius, r$_{bulge}$, where the two intersect. 
Within this cut the mass contribution of the disk stars is much smaller than the spheroid, 
therefore minimizing the contamination from disk stars when determining the bulge global properties.
An example of projected stellar mass density for the disk, spheroid and galaxy stellar particles is given in Figure~\ref{mass_dens} for G-1536, as well as the 
position of the radial cut we use to define the bulge region.

Although a dynamical separation of galaxy components is of great help in analysis of galaxy structure formation, 
from an observational point of view the different dynamical components of the central galactic regions are usually hard to disentangle. 
For this reason, we also use an observationally based selection for the bulge stars, in the form of a simple radial cut at r$_{bulge}$.

Summing up, we use two bulge selection criteria: i) the stars within an sphere of radius $r_{bulge}$, hereafter position selection, 
and ii) the stars within an sphere of radius $r_{bulge}$ that at the same time belong to the galaxy spheroid according to the kinematic-decomposition, hereafter kinematic selection. 
In Table~\ref{tab1} we give $r_{bulge}$, the total stellar and gas mass of each bulge at $z=0$, as well as the mass resolution for each simulation.

\section{Bulge populations}
\label{stepops}

\begin{figure*}[t!]
\centering
\includegraphics[scale=.85]{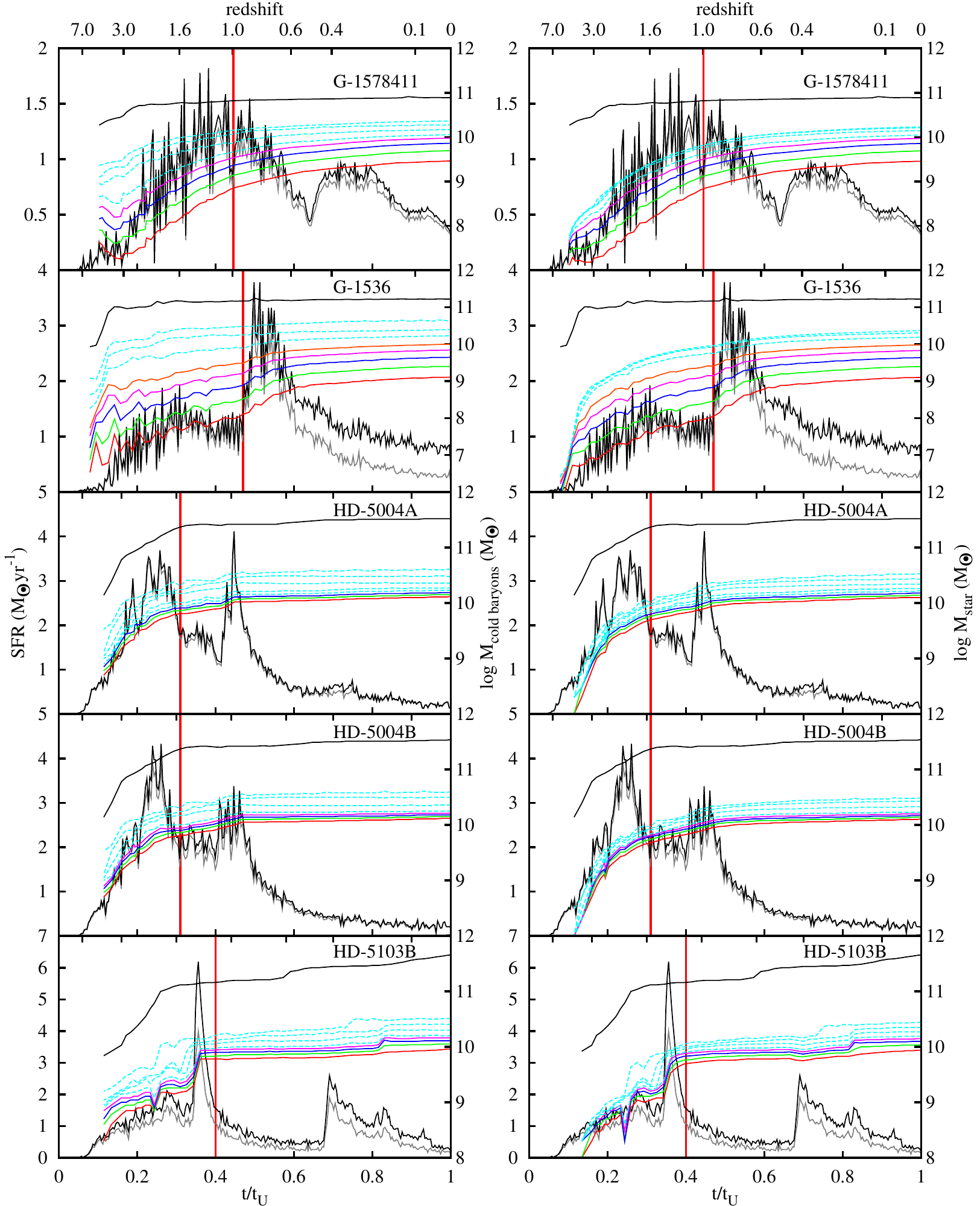}
\caption{
Bulge star formation rate comparisons for kinematic (solid gray) and position (solid black) selected bulge stars with sizes given by r$_{bulge}$. The position of the thick red vertical 
(values in Table~\ref{tab2}) indicate the separation between the old and young stellar populations. 
The colored lines represent the mass aggregation tracks (MATs)  of cold baryons (left) and stars (right) 
along the main branch of the merger tree for each object, the masses being computed inside fixed radii of 0.50, 0.72, 1.03, 1.47, 2.10, 4.30, 8.80 and 25.80 kpc from bottom to top. 
The upper solid black line is the total mass inside the virial radius (virial mass), the solid curves in red, green, blue, magenta and orange correspond to radii roughly within the bulge, 
while the dashed cyan ones correspond to radii exceeding $r_{bulge}$ (see Table~\ref{tab1}).}
\label{sfr-mass}
\end{figure*}

To analyze the bulge stellar populations and their link with dynamical processes, we constructed Figure~\ref{sfr-mass},
in which both the star formation rate (SFR) histories of the bulge  and the galaxy mass aggregation tracks (MATs) have been drawn. 
The MATs each give the  evolution of mass inside a fix radius (right axis in each panel). Virial radii have been calculated based upon the \cite{Bryan:1998} fitting function 
to determine the over-density threshold. For the baryonic component, radii are binned in equally spaced steps in a logarithmic scale. 
On the right (left) panels the colored curves show the stellar (stellar plus cold gas) masses within the respective radii as a function of the Universe age 
(t$_U$ is the Universe age at $z=0$) and redshift.
MATs corresponding to radii within $r_{bulge}$ are drawn with different colors, while those at larger radii are given in cyan.

Major mergers ($M_{secondary}/M_{main} > 0.25$), minor mergers and 
slow accretion processes in the dark matter or baryonic component can be clearly identified as big or small mass jumps in the MATs, or as continuous mass increments, 
respectively (note the different mass scales on the right axes for the {\tt P-DEVA} and {\tt GASOLINE} galaxies). 
Two different phases are reflected in the noticeable knee-like shape of the MATs in all objects shown in Figure~\ref{sfr-mass}:  
 an initial phase where the mass assembly rate is high, and then a much slower phase, when mass is more slowly acquired. 

Superimposed on the same figure are the bulge SFR histories (left axis in each panel), for both position and kinematic selected bulge populations. 
All objects present starbursts at large redshifts, peaking between $z = 3$ and $z = 1$,
where  significant  \textit{substructure} is also apparent in many cases. 
These early star formation  peaks are expected when the primordial gas suffers a violent collapse and are  correlated with the knee-like feature visible in their corresponding MATs. 
The high redshift SF peak is a direct consequence of the first phase of mass assembly. 
The simulations also show later peaks in star formation that are related to  merger/accretion events. 

At lower redshifts a variety of processes appear imprinted in the SFR histories: 
during quiescent periods, the SFR decays to an approximately constant tail at the same time as the MATs get flattened. The low tails present in all SFR histories correspond 
to periods of mass assembly where only minor mergers or gas accretion at slow rates show up in the MATs. All simulation also show secondary jumps in SFRs, associated 
with later merger events. These vary in time, and in size relative to the initial starbursts, for example HD-5103B has a relatively late (t/t$_U$) accretion event 
that shows up in the total mass MAT (black line) as well as the build up of bulge mass and the star formation rate. 

Some qualitative differences are apparent in the two {\tt GASOLINE} runs 
compared to the {\tt P-DEVA} ones. The initial starbursts associated with the fast phase of mass accretion are less peaked in the former, due to the affects of feedback. 
In  G-1578411, which has the  lower mass but also lower  feedback  of the two {\tt GASOLINE} simulations, the initial starburst has a longer duration than the other simulations, 
while in G-1536 the initial starburst in the bulge region is more suppressed relative to the overall star formation within the bulge, due to the more efficient feedback in this simulation. 
Also, due to energetic feedback, the MATs of  G-1578411 and G-1536 show delayed baryon mass assembly relative to their halos collapse at high $z$s.
As a result, these objects have low central baryonic mass density, the effect being more marked at high-$z$.

Regardless of these specific differences, the correlations between MATs and SFR histories suggest that a meaningful stellar age classification can be based upon the two phases showing up in the MATs. 
Therefore we classify as old bulge stellar population the stars formed as a direct consequence of the fast phase 
(with formation times smaller than the temporal cut drawn in thick red in Figure~\ref{sfr-mass}), while the stars formed later on we globally denote by young bulge stellar populations.
In a quantitative sense, the temporal cuts correspond to the time from which the second derivative of the MAT(r$_{bulge}$) becomes flat, reflecting a transition from a
fast clumpy mass assembly to a slow smooth regime, without considering the MAT variations induced by low redshift mergers, like in the case of HD-5103B.  
The values of the temporal cuts, t$_{cut}$, for all bulges are given in Table~\ref{tab2}, together with the mass weight of the old population.

\begin{table*}
\begin{minipage}{6.3in}
\renewcommand{\thefootnote}{\thempfootnote}
\begin{center}
\begin{tabular}{|c|c|c|c|c|c|c|} \hline
Object & t$_{cut}$/t$_U$ & z$_{cut}$ & M$_{old*}$/M$_*$& n$_{old}$ & n$_{young}$ & n$_{total}$\\ \hline\hline
G-1578411 & 0.45 & 0.98 & 0.39 & 1.17 & 1.07 & 1.22\\
G-1536 & 0.47 & 0.90 & 0.31 & 1.19 & 1.59 & 1.39\\
HD-5004A & 0.31 & 1.54 & 0.43 & 2.95 & 2.09 & 2.57\\
HD-5004B & 0.31 & 1.54 & 0.42 & 3.96 & 3.38 & 3.23\\ 
HD-5103B & 0.40 & 1.12 & 0.50 & 3.49 & 3.27 & 3.68\\ \hline
\end{tabular}\\
\caption
{The temporal separation between the distinct star formation episodes for the simulated bulges,the corresponding mass percentage of old and young stars in the position-selection,
and the S\'{e}rsic indices derived from fitting the projected stellar mass density of old and young (according to the bulge temporal cut), and of all galaxy stellar particles.} 
\label{tab2}
\end{center}
\end{minipage}
\end{table*}

\section{3D Shapes, sizes and kinematics}\label{shape}

\begin{figure}[h!]
\includegraphics[scale=1.0]{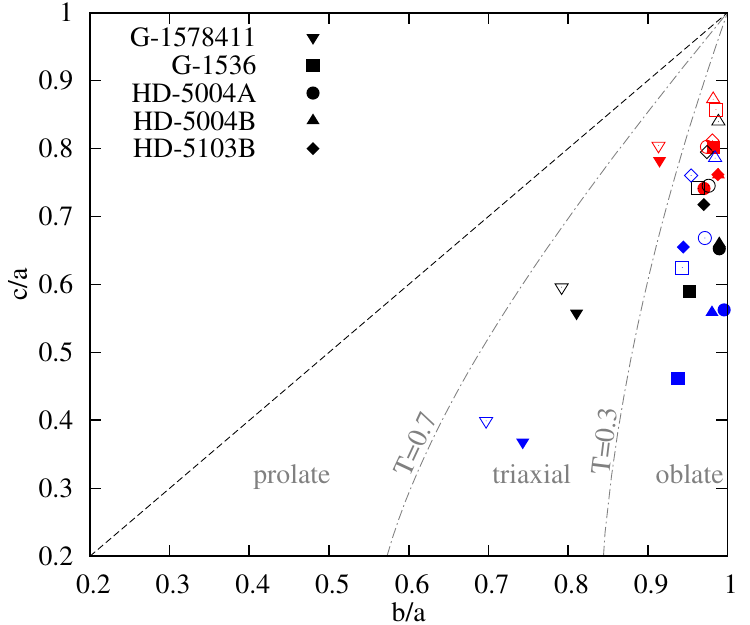}
\caption{The axis ratios of the five spheroids. Red and blue represent the old and young bulge, while black gives the average over the whole object.
Filled and empty symbols correspond to the objects in the position and kinematic selection, respectively.
The dashed-dotted gray curves separate prolate, triaxial and  oblate morphologies.}
\label{axisratios}
\end{figure}

To quantify the shape of the bulges, as well that of  the old and young bulges as defined above,  we show the correlations between the axis ratios 
of each ellipsoid of inertia in Figure~\ref{axisratios}. The inertia tensor was computed following \cite{Gonzalez:2005}, and subsequently diagonalized. Next, the eigenvalues 
($\lambda_1$~$>$~$\lambda_2$~$>$~$\lambda_3$) were used to compute the length of the principal axis a~$\geqslant$~b~$\geqslant$~c.  
Red and blue represent the old and young bulge, while black gives the average over the whole object.
Filled and empty symbols correspond to the objects in the position and kinematic selection, respectively.

\begin{figure}[h!]
\includegraphics[scale=1.0]{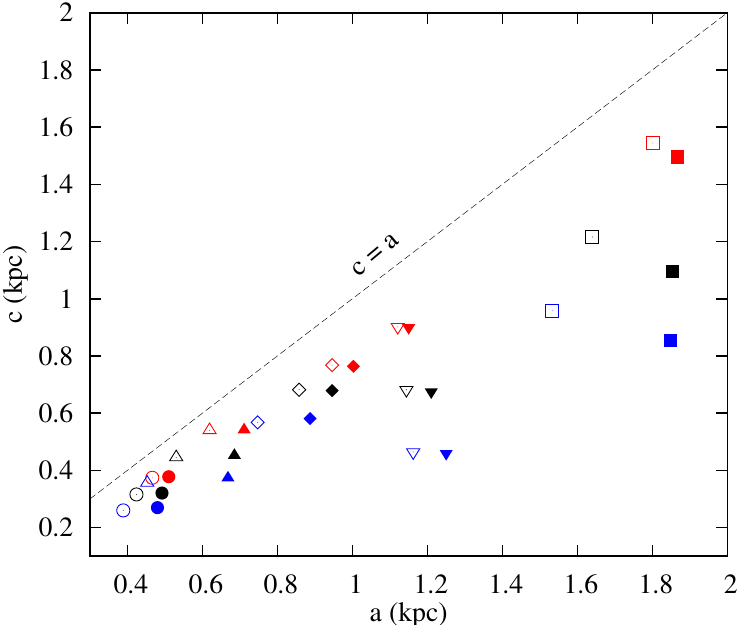}
\caption{Minor vs major semi axis for the five spheroids.   Red and blue represent the old and young bulge, while black gives the average over the whole object.
Filled and empty symbols correspond to the objects in the position and kinematic selection, respectively. The dashed line is where c=a.  
Young bulge populations are smaller than old ones, with the exception of G-1578411.  }
\label{majmin}
\end{figure}

\begin{figure}[h!]
\includegraphics[scale=1.0]{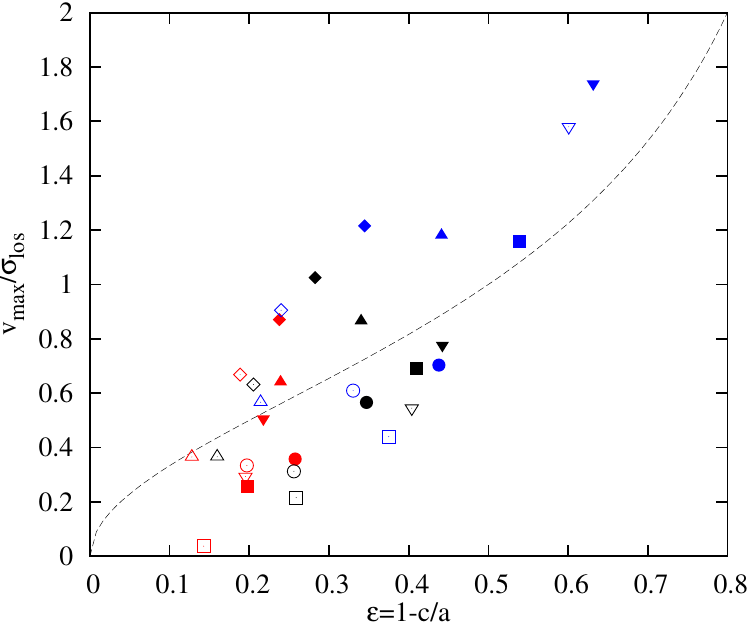}
\caption{Rotational support versus ellipticity.  Red and blue represent the old and young bulge, while black gives the average over the whole object.
Filled and empty symbols correspond to the objects in the position and kinematic selection, respectively. 
Young bulges are more rotationally supported and less spherical than the old populations, even when kinematically selected to exclude the disk. }
\label{rotsupport}
\end{figure}

The objects situated on the dashed black line in Figure~\ref{axisratios} have $b=c$, while the dashed-dotted gray curves denoted by $T=0.3$ and $T=0.7$, respectively, separate the oblate objects 
from the triaxial and the prolate ones according to the $T$ parameter introduced by \cite{deZeeuw:1991} with the definition $T=(1-(b/a)^2)/(1-(c/a)^2)$. 
The oblate objects correspond to $c/a<0.9$ and $T<0.3$, the prolate to $c/a<0.9$ and $T>0.7$, and the rest to triaxial ones. 
In this perspective, we observe that both complete bulges as well as their distinct components are all oblate with the exception of 
the secular bulge of  G-1578411 which is triaxial regardless of the selection criteria. 
Old and young populations are clearly segregated, either in the position or in the kinematic selections, and more separated in $c/a$ than in $b/a$. 
Old populations are more spheroid-like, while the young are more oblate,  consistent with observations. 

 Figure~\ref{majmin} gives an idea of the sizes of the five bulges by plotting $c$ vs $a$. In all but one case (the $a$ value of G-1578411), 
the old bulges have larger sizes (as given by $a$ and even more by $c$) than the young ones, sustaining the idea that younger 
stellar populations ocupy a smaller volume than their corresponding old counterparts. In this respect the following inequality holds, irrespective
of the selection criteria used: $\sqrt[3]{a_{y}b_{y}c_{y}} < \sqrt[3]{a_{t}b_{t}c_{t}} < \sqrt[3]{a_{o}b_{o}c_{o}}$, where $y$, $t$ and $o$ stand for young, total and old, respectively.
This graph shows a clear sequence, the size increasing from the younger to the older, with the total 
population in between. Also, as it was proven by the previous graph, the older bulge components are closer to be perfect spheroids (they appear closer to the dashed line for which $c=a$).

The difference between the results obtained using the two selection criteria are  only qualitative.
 Indeed, the trends in shapes and sizes are the same for both selections.
The kinematic-selection leads, on average, to bulge components closer to being spherically symmetric than the position selection does. 
This is an expected result since position selection implies the exclusion of precisely some of the particles encountered in ordered rotation 
which would otherwise lead to ellipsoids of inertia characteristic of more disk-like objects. 

Also, the shape and size trends with stellar age are the same for the {\tt P-DEVA} and {\tt GASOLINE} bulges. 
The only difference is that {\tt GASOLINE} bulges have lower mass concentrations and therefore larger values for $a$ and $c$, due to energetic feedback.

Let us now turn to the bulges kinematical analysis, including some of their classical or pseudo-bulge properties,
therefore using line-of-sight (los) velocities.
In order to analyze the rotational support, we align the galaxies with the $z$-axis perpendicular
to the disk. Since the rotational velocity can be best observationally measured in edge-on projection, we
consider two lines of sight (along the $x$- and $y$-axes), and compute the radial profiles of the velocities -- $v_{\rm los=x}(R=y,z)$ and 
$v_{\rm los=y}(R=x,z)$, respectively -- along the two. The dependence of the profiles on the altitude $z$ above the galactic plane is important both when studying whether the bulge
or its components are in cylindrical rotation, as well as from the point of view of comparison with observations. 
In the latter case, data are normally taken, due to the extinction effect in the disk plane, with long-slit or IFUs at altitudes higher than the vertical scale height of the thin disk.
Considering all these observational limitations, we constructed the velocity profiles using a $z$-binning of 0.2 kpc, 
averaging the curves in the two hemispheres in order to minimize the statistical noise in the slits at higher latitudes where the number of particles is smaller. 
Complementary to the profiles of rotational velocity, we also constructed the los velocity dispersion radial profiles, 
$\sigma_{\rm los}(R,z)$, which are approximately flat. 
Given the flatness of these profiles, we considered the $\sigma_{\rm los}(z)$ as the average of $\sigma_{\rm los=x \mid y}(R=y \mid x,z)$ over all lines 
of sight $R=y \mid x$\footnote{$y \mid x$ stands for $y$ or $x$}, 
and weighted with the number of particles in each bin $n_{R}$. Once the rotational velocity profiles were constructed, we took as $v_{max}(z)$ the weighted average of the profile extremes. 
Finally, we averaged $v_{max}(z)$ and $\sigma_{\rm los}(z)$ over the two lines of sight, los=$x$ and los=$y$.

In order to choose a representative $z$-bin for all bulges, we considered both the disk vertical scale heights as well as the bulge radial extensions. 
Therefore, we selected the bin with $0.4 < \mid z \mid < 0.6$ kpc. Given this choice, within the current section, $v_{max}$ and $\sigma_{\rm los}$ will refer to this specific slit.

 Figure~\ref{rotsupport} plots rotational support, defined as the maximal 
rotational velocity ($v_{max}$) divided by the bulge los velocity dispersion ($\sigma_{\rm los}$), 
as a function of the intrinsic ellipticity $\varepsilon = 1-c/a$.
The dashed black curve in the graph represents oblate-spheroid systems with isotropic velocity dispersion flattened only by rotation \citep{Binney:1978}, 
and is approximately described by $\sqrt{\varepsilon/(1-\varepsilon)}$ \citep{Kormendy:1982}.  

The results show a clear trend of higher eccentricities and rotational support for the younger populations (blue symbols) as compared 
to the older ones (red symbols), while the whole bulges, in black, occupy intermediate positions. 
It is important to note that these patterns hold irrespective
of the code used to run the simulations or the bulge selection criteria, as well as of the slit positioning. 
Note however that increasing (decreasing) the $z$ slit positioning lowers (increases) the specific values of $v_{max}/\sigma_{\rm los}$ for
all the five bulges as well as in their distinct stellar components. Therefore, our bulges appear more classical as we move away 
from the galactic plane, but always they show same trends with stellar age, with the younger component being more rotationally supported. 
We note that in all cases, a weighting by light rather than mass will provide increased prominence to  the younger pseudo-bulge-like  populations.
In two cases (HD-5004A and HD-5103B in the dynamical selection), the bulge as a whole has a slightly lower value of $v_{max}/\sigma_{\rm los}$ even than the old population.
In these cases, the bulges have velocity dispersions approximately equal to the old components, but have slightly smaller $v_{max}$. This difference in $v_{max}$ can be explained through 
misalignments and/or kinematical peculiarities.

\begin{figure*}
\centering
\includegraphics[scale=1.0]{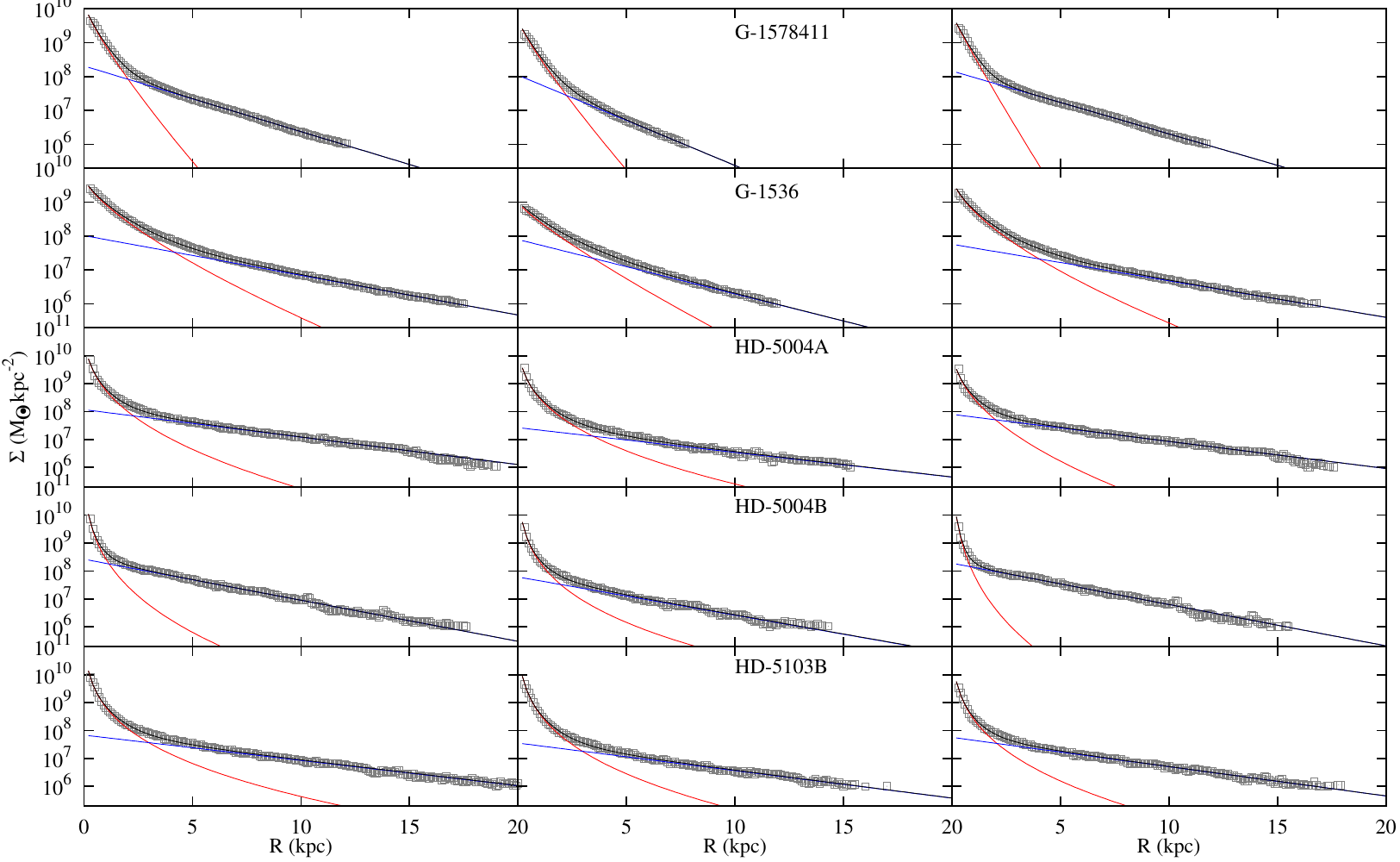}
\caption{The S\'{e}rsic + exponential disk fits of the projected mass density of galaxy stars (left panels), and of the corresponding old (central panels) and young (right panels) stellar sub-populations, 
separated according to the bulge temporal cut in Table~\ref{tab2}. The color code is: grey open squares for the data, red and blue lines for the S\'{e}rsic and exponential disk contributions,
and black lines for the total fit.}
\label{sersic}
\end{figure*}

Another parameter describing the bulge shape is the S\'{e}rsic index ($n$).
Thus, we fitted the projected stellar mass density profiles of our five galaxies with a S\'{e}rsic + an exponential disk. The data and the fits are depicted in 
Figure~\ref{sersic}, while the S\'{e}rsic indices are given in Table~\ref{tab2}. With the aim of checking whether $n$ varies if considering only the old or only the
young stars, we also fitted the two stellar subpopulation of the galaxies, using for the temporal cut the limit defined with respect to the bulge region. It is important 
to stress, that $n$ values are to a large extent dependent on the amount of feedback. In the case of {\tt GASOLINE} galaxies, the incresed feedback leads to a flattening 
of the projected stellar mass curves at small radii. For this reason, the S\'{e}rsic indices of G-1578411 an G-1536 are $< 2$ irrespective of considering all
galaxy stars or only the corresponding components, while those of HD-5004A, HD-5004B and HD-5103B (simulated without explicit energetic feedback) are $> 2$. 
In any case, the difference between $n_{old}$ and $n_{young}$ is small in any of the simulated galaxies, with a slight tendency for the old stars to have larger $n$ 
than the younger ones, with the exception of G-1536, which was simulated with a considerable amount of feedback. From an observational point of view, no define trend with the band  
has been detected either, see for example \citet{Fisher:2008}, or more recently \citet{McDonald:2011}. 
The S\'{e}rsic index has been used, together with other parameters like rotational support, to classify bulges. 
For example, \cite{Kormendy:2004} consider pseudo bulges to have $< 2$, while classical ones
have $> 2$. In this respect, the three {\tt P-DEVA} galaxies have classical bulges, the old bulges being more classical than the young ones ($n_{old}>n_{young}$). On the other hand, 
the two {\tt GASOLINE} ones have pseudo bulges according to this classification, and show no definite trend with age. In the same perspective, the pseudo bulges of \citet{Okamoto:2012},
formed at high redshift, roughly during what we define as the fast phase of mass assembly, have $< 2$. However, his simulations also use energetic feedback which is at least partially 
responsible for the low values of $n$. Therefore, we think that further studies are needed in order to disentangle to what extent the $n$ values depend on the particular modes of mass assembly
and to what extent on the amount and implementation of the feedback effects. In this respect, our study suggests that, at least for simulations, $v/\sigma$ separates more reliably 
the young from the old bulge than $n$, which tends to be similar for old and young stellar populations.

\section{Ages and abundances}\label{chem}

This section concerns the ages and metallicities of the old and young bulge stars at $z=0$.

\begin{figure*}
\centering
\includegraphics[scale=.85]{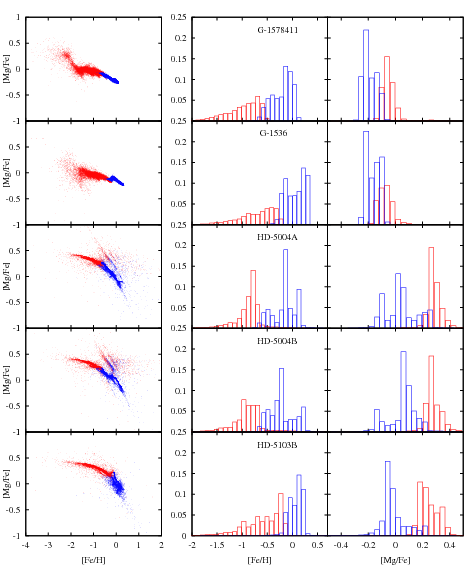}
\caption{
[Mg/Fe] vs [Fe/H] (left), and the [Fe/H] (center) and [Mg/Fe] (right) histograms for the simulated bulges. 
The bin widths for [Fe/H] and [Mg/Fe] are 0.07 and 0.04 dex, respectively. Red and blue represent the old and young bulges.}
\label{chemcomp}
\end{figure*} 

Although details differ slightly, in  both the {\tt P-DEVA} and {\tt GASOLINE} simulations trace the production and enrichment of chemical elements in broadly the same way. 
Newly produced elements, as by-products of stellar evolution and death, are released to the surrounding interstellar medium as increments in the metal content of nearby gas particles. 
Metal diffusion is implemented among gas particles with a diffusion constant, 
allowing for elements to mix in a given environment, mimicking what happens in a turbulent interstellar medium. 
Both codes consider the evolution of the following elements: H, He, C, N, O, Ne, Mg, Si, S, Ca and Fe. 
As we are looking at broad trends here, the relatively small differences in the implementation chemistry between the codes  are not critical. 
As a measure of metallicity we used either [Fe/H] or the average values of these elements; 
also, we used [Mg/Fe] to follow $\alpha$-element properties and checked that using [O/Fe] does not change substantially our results. 
The reference values for the solar metallicities were taken from \cite{Grevesse:1998}.  

We analyze the distributions of metallicity and $\alpha$-element abundances in the different bulge populations. 
They are plotted in Figure~\ref{chemcomp} (central and right panels), 
where it can be seen that the abundance distributions of young and old bulge populations are  segregated (the old stars have a lower metallicity and a higher [Mg/Fe]), 
giving in most cases overall two-peak [Fe/H] and [Mg/Fe] distributions. 
The slightly cleaner separation of two populations in [Mg/Fe] space in the {\tt P-DEVA} simulations as compared to the {\tt GASOLINE} ones 
might partially be an effect of the energetic feedback in the latter,
which delays the SF relatively to dynamical processes (see Figure~\ref{sfr-mass} and the corresponding comments in Section~\ref{stepops}) and enhances the  mixing, 
although  the differences in the details of implementation of metal enrichment could also play a role,
as well as in the wider ranges of [Mg/Fe] of the P-DEVA runs.  However, the important point here is that both codes result in broadly the same 
relative trends in metal abundances and enrichments of their old and young populations.

In this respect, we also plot [Mg/Fe] as a function of [Fe/H] (left panel). Again the young and old populations show up at different loci in this plot.
A feature to be noted in the graphs [Mg/Fe] versus [Fe/H] are the slopes of the two stellar population types in the sense that the old 
one has a milder slope, while the young population shows a steeper one. 
Again, these trends are apparent in all simulations, and what is interesting is that the age selection we have used invariably separate at the "knees" of the abundance trends. 
This feature has to do with the different nucleosynthetic origin of $\alpha$ and Fe elements,
a clean knee-like pattern being expected in stellar populations 
where the SFR is concentrated in an episode with a very short time-scale followed by an epoch of more quiet star formation. 
The plateau corresponds to the early stages when SNe II dominate the metal 
enrichment, followed by a downturn to lower [Mg/Fe] when enough time has elapsed since the high redshift starburst that SNe Ia come into play 
\citep[see][and references therein]{Wyse:1999}.

These results on the chemical composition of bulge populations agree with recent observations of the Milky Way Bulge, like those of 
\cite{Babusiaux:2010}, 
who found two distinct populations in Baade's Window, one old, metal-poor component with kinematics of a spheroid and a younger, metal-rich one displaying kinematic 
characteristics similar to a bar. Otherwise, \cite{Ness:2012} extended this work by enlarging the number 
of windows, finding consistent results (see Section~\ref{intro} for more details).

\begin{figure}
 \centering
  \includegraphics[]{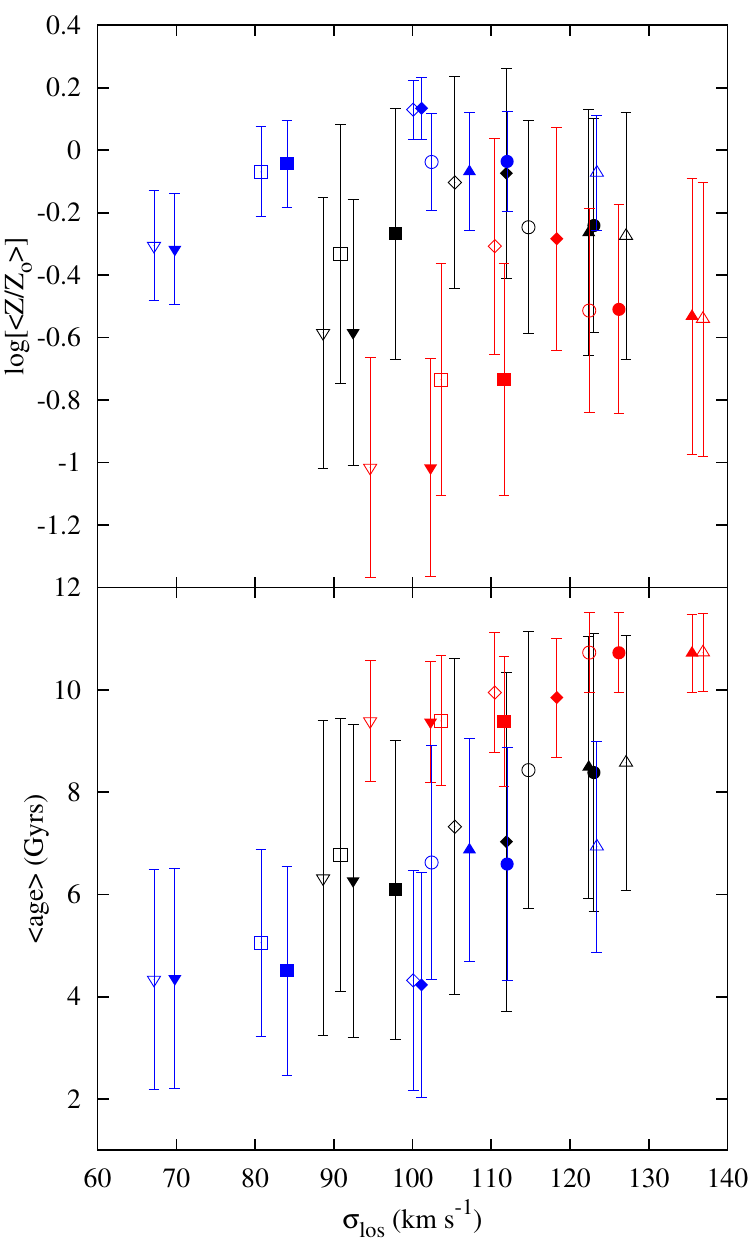}
  \caption{Average metallicity (top) and age (bottom) of the distinct bulge components versus los velocity dispersion. 
Red and blue represent the old and young bulge, while black gives the average over the whole object.
Filled and empty symbols correspond to the objects in the position and kinematic selection, respectively.}
\label{age_met}
\end{figure}

Figure~\ref{age_met} gives the averaged metallicities and ages of the distinct stellar populations as functions of the los velocity dispersion. 
In this case, $\sigma_{\rm los}$ was computed in the same way as in Section~\ref{shape}, the only difference being that no $z$-binning was used.
First of all, we note that a clear population segregation stands out in both panels, irrespective of the code or the selection criteria employed. 
In the upper panel, the metallicity appears to be increasing with the velocity dispersion, when considering all five bulges. However, if discarding 
the  G-1578411 bulge (bottom-up triangles), $\left\langle Z \right\rangle$ seems to be almost constant with $\sigma_{\rm los}$. On the other hand,
if considering only the {\tt P-DEVA} bulges, $\left\langle Z \right\rangle$ decreases with $\sigma_{\rm los}$. 
These trends should be considered with caution,
given the inhomogeneity of our sample.
   In the age -- $\sigma_{\rm los}$ plot, we get a positive slope, in accordance with the findings of \cite{MacArthur:2009}. 
Taking $\sigma_{\rm los}$ as a measure of the objects mass \citep{Binney:1987} (see Table~\ref{tab1} for the bulge stellar masses), it can be noted 
that the more massive objects also have the older average overall populations.

\section{Bulges within The Cosmic-Web}
\label{twophase}

\begin{figure*}
\centering
\includegraphics[scale=.80]{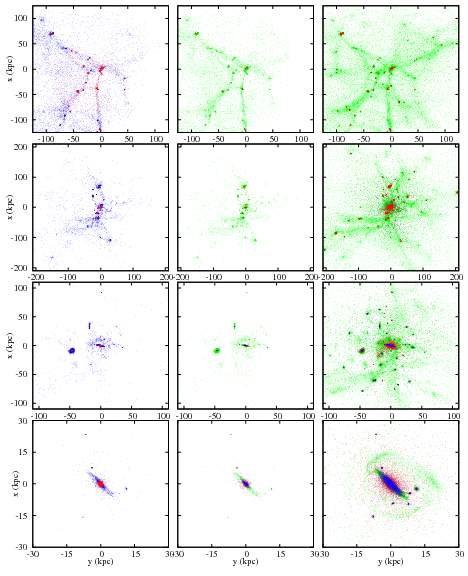}
\caption{The different components of the bulge HD-5004A (left and central panels) and its surroundings (right panels) at redshifts 3.50, 1.75, 1.21 \& 0.58 ($t/t_U$ of 0.13, 0.28, 0.38 \& 0.59),
from top to bottom. The left panels give the positions of the bulge stellar particles identified at $z=0$ when traced back to each of the four $z$s. 
Red corresponds to baryon particles (either gas or stars) whose transformation into stars occurs along the fast phase of bulge formation (i.e., at  $t_{\rm form}/t_U < 0.31$, or equivalently, 
$z > 1.54$ in this particular case),
 and blue to bulge baryon particles that become stars at later times. The same traced back positions are given in the central panels, 
but in this case with a color code representing particles properties {\it at the
$z$s plotted}, namely green for cold gaseous particles ($T < 10^6 $K), black for the hot ones, and red and blue for the stellar particles according to their time of birth (older and respectively 
younger than $z = 1.54$). The right panels show the positions of all baryonic particles within a limiting radius of the galaxy's progenitor center using the same colors as in the central graphs.}
\label{snapshot}
\end{figure*}

We trace the evolution of particles in the bulge at 
low $z_{low}=0$ to their origins at $z_{in}=10$, following them at 60 different redshifts in between.
In Figure~\ref{snapshot} we plot the evolution of the baryons that form the bulge stars of HD-5004A at $z=0$ (left and central graphs) at four $z$s (top to bottom) 
representing relevant consecutive events in this bulge formation,
 as well as the object's surroundings (right graphs) at each redshift\footnote{Indeed as Figure~\ref{sfr-mass} shows, the first snapshot
at $t/t_U = 0.13$ gives us the configuration of the bulge-to-be stars at the very beginning of 
the collapse-like event;
the second one at $t/t_U = 0.28$ roughly corresponds at the end of the fast phase; the third one at
$t/t_U = 0.38$ is just before the beginning of the  major merger 
causing the strong stellar burst in the slow phase,
and finally the fourth one at $t/t_U = 0.59$ corresponds to the end of this burst.}.
In the other simulations this sequence of evolution is similar enough that we can use this as a representative example.   
The left sequence aims at illustrating the differences in the space configurations that old $z=0$ bulge-to-be  stars (red points) 
form along their evolution as compared to those of young $z=0$  bulge-to-be stars (blue). 
The emphasis here is on the two population types, irrespective whether they are gas or stars at the Universe age corresponding to the plot. 
The central sequence takes into consideration the bulge-to-be stellar particle properties at the $z$s plotted, 
namely points are  green for cold gaseous particles ($T(z)<10^6K$), black for hot ones ($T(z)>10^6K$), 
and red and blue for stellar particles, depending  on their  time of birth  according to Table~\ref{tab2}. 
Its aim is to illustrate the differences in the birthplaces of the two bulge stellar populations relative either
to the cosmic-web (at high $z$s) or to the protogalaxy components (at lower $z$s),
as well as their evolution towards their final configuration at $z=0$.  Finally, in the right panels
we plot all the particles in the same volumes, and with the same color code as in the central one, 
in order to compare the dynamical processes of the bulge-to-be stellar particles with those of the surroundings mass elements causing them.

 We first analyze the sequence of bulge formation events in terms of the cosmic-web element dynamics (central panels).
Before  $z=3.5$  mass piles up in caustics \footnote{ Caustics are the elements of the cellular structure
where dense  mass elements show up at high $z$s. They  are classified into walls, filaments and nodes,
and, at a given scale, these represent a temporal sequence of mass piling up. See \citet{Shandarin:1989} and \citet{DT:2011} for more details.},
 where already some nodes show up. 
At this redshift, star formation is already triggered in the densest of these nodes at {\it separated places}.
Comparing the two upper snapshots in any of the vertical sequences, 
we see that an overall collapse-like event acts in between onto a structured net of cells as a  contractive deformation. 
It somehow erases the cell structure, joining together clumps mainly along filaments, and therefore
only very low relative angular momentum is involved at these $z$s. 
The collapse-like event   shrinks the volume visualized at $z=3.5$ into that at $z=1.75$,
where a central mass concentration fed by filaments stands out. Gas keeps on flowing through filaments towards
the central regions between snapshots two and three.
 A diffuse component (i.e., outside caustics) is also evident at high $z$. 
This diffuse component tends to flow into caustics, therefore vanishing as evolution proceeds. 
Otherwise, by  $z=1.21$ the filaments have 
practically been removed in favor of clumps and some cold gas in irregular structures. Moreover, 
 two small gaseous disk-like structures 
and a third central disk with a stellar component, seen edge-on, have formed in the interval
between the second and third snapshots. 
At $z=1.21$, the stars are at the center of these disks, 
or at the center of other smaller gaseous structures. The young-to-be bulge stars begin to show up (blue points).
Finally, after a  major merger occurring between the third and fourth snapshots,
by $z=0.58$ practically all the baryonic particles that are to form the HD-5004A bulge stars at $z=0$ come to be bound in a unique system (except for some small clumps). 
Most of the particles that have transformed into stars (red and blue at the central sequence) are at the center, 
while those that are still gaseous (in green) show up in a rebuilt disk-like configuration. This configuration 
disappears later on, and by $z=0.35$ (not shown in Figure~\ref{snapshot}) practically all the bulge particles are at their place within 
a sphere of radius r$_{bulge}$.

It is very useful to compare these evolutionary processes with the mass distribution of their surroundings visualized in the right graphs of Figure~\ref{snapshot}.
This comparison makes it even clearer that bulge material has been involved in caustic formation at high $z$ although part of it remains as diffuse gas quite a while.
Interestingly, a fraction of it has formed transient gaseous disks or has been a part of the former disk of the HD-5004A object. 

We now  analyze the differences in the assembly patterns of the young and old bulge stellar populations relative to the cosmic-web dynamics (left graphs in Figure~\ref{snapshot}). 
Most mass elements identified at $z=0$ as old stellar particles (red points) are already concentrated in caustics by $z=3.5$. 
These  are the densest regions  at this time. In contrast, an important fraction of the mass elements identified at $z=0$ as young stellar particles 
(blue points) are not yet involved in caustics. The difference increases as evolution proceeds. In fact, 
the strong contractive deformation acting between the first and second snapshots involves preferentially
the old-stellar-to-be particles, in such a way that
by $z=1.75$ the old component has a very small volume, while the young component still spans a scale of $\sim200$~kpc and has
an important diffuse gas fraction (with some nodes formed at local contractions as well).
In the third snapshot we see that a fraction of the mass elements that are to form young stars now shows disk-like patterns, while old stars are at the centers of these small disks. 
These disk-like configurations provide a key to explain the different kinematic properties of young bulge populations as compared to old ones, 
because of the angular momentum content of the former.
As said above, the disks seen in the third snapshots merge.  This causes their central old stars to
form a unique bulge-like structure. At the same time, 
new stars -- belonging to the young bulge --  form and appear within it. On its turn, the young component, partially still as diffuse gas, forms structures
(i.e. a disk in the case of HD-5004A as it can be seen in Figure~\ref{snapshot}) around the old stars spheroid.
 Some of  the small nodes remain as 
satellites, carrying a small bit of old population at their centers.
Finally, by $z=0.35$ (not shown in Figure~\ref{snapshot}), both the old and young bulge populations (or the gaseous particles to form the latter) are bound to their respective central spheroids, 
where they remain until $z=0$.  

We draw attention to similarities with massive ellipticals \cite{DT:2011}. 
In this case, stellar particles at $z=0$ exhibit a gaseous web-like morphology at $z\sim3.5\div6$, with scales of $\sim1$ physical Mpc.
The densest mass elements of this gaseous web, dynamically organized as  
attraction basins for mass flows, have already turned into stars by $z\sim6$. 
At high $z$ these basins undergo fast contractive deformations which violently shrinks them in quasi-radial
directions, therefore involving very low angular momentum. 
They can be described as collapse events with very complex geometries, causing high rates of dissipation
and stellar formation out of the available gas, most of if it being transformed into stars at the end of
this process. Afterwards, during the second phase, the mass assembly rate is much lower and is characterized 
by mergers involving significantly larger amounts of angular momentum.

To sum up, the old and young populations of the bulge HD-5004A are not only segregated in their $z$=0 properties, 
but also show quite different assembly patterns in terms of the cosmic-web dynamics. 
Overall, this scenario also holds for the other four bulges.  
Indeed,  most of the old stars formed at disjoint places within attraction
basins during the early fast processes, and undergo an important assembly episode later on, through
multiclump collapse-like events shrinking these basins in quasi-radial directions.
This assembly phase results in a relative low  angular momentum of the merging clumps.

The ancestors of young bulge stars, on their turn, in many cases have been part of disk structures before arriving to
the galaxy central regions. Different destabilizing processes can be responsible for this inward
material transport, such as mergers or secular processes in the disks. Young bulge stars keep 
partial dynamical memory of their angular momentum content.
These differences in the assembly patterns relative to the angular momentum involved,
would be a key piece in explaining the segregated properties of the two bulge components.
Although our sample is small, the mass weight of the old component varies between 31 and 50\%. 
An even more important variation  of this weight, together with the different
inward mass transport mechanisms in the second assembly phase, could explain the important dispersion
in bulge properties.

\section{Summary, Discussion and Conclusions}
\label{summary}

We have analyzed the bulges of an inhomogeneous suite of five  spiral galaxies emerging from high resolution hydrodynamical simulations in a cosmological context.
Our aim is to decipher the underlying physical processes in bulge assembly that could be responsible for their observed properties,
focusing on relevant structural, kinematical and chemical  properties of their stellar populations 
and paying particular attention to their observationally suggested segregation by age.

Concerning the problem of bulge formation, the main particularity of this work is
that we analyze simulated disk galaxies that are run with different codes (including different sub-grid physical
prescriptions) where the chemical evolution has been carefully implemented, 
by using a common pipeline to measure the relevant stellar population properties.
 The simulations have been run with {\tt P-DEVA} and {\tt GASOLINE}.
Feedback from massive stars is implemented explicitly in the  two  {\tt GASOLINE} simulations
through the blast-wave formulation \citep{Stinson:2006}, while in {\tt P-DEVA} simulations the effects of discrete
energy injection are assumed to be on subgrid scales, resulting in the low star formation efficiency which is assumed
to mimic them \citep[see  discussion in][]{Agertz:2011}.

Although details differ slightly, both {\tt P-DEVA} and {\tt GASOLINE} simulations trace the production and enrichment of chemical elements in broadly the same way. 
The newly formed elements, as by-products of stellar evolution and death, are released to the surrounding interstellar medium as increments in the metal content of nearby gas particles. 
The metals are diffused among gas particles, allowing the elements to get mixed and therefore, mimicking the turbulent interstellar medium.
Both codes consider the evolution of 11 elements. 
As we are looking for broad trends here, the relatively small differences in the implementation chemistry between the codes  is not critical.

In this paper we have studied in detail the mass-averaged three-dimensional sizes, shapes and kinematics, as well as stellar ages and element compositions of these five bulges.
This is an intrinsic approach as opposed to looking for quantities closer to observations (i.e., light-averaged).
We keep as this level because our aim in this paper is to find out the patterns of bulge mass assembly by focusing on their stellar population properties.

We have found a satisfactory qualitative agreement with the latest observational data for the Milky Way as well as for other external bulges (see Section~\ref{intro}).
Our results indicate that bulges in our sample have an old stellar population, 
formed at high $z$ at disjoint places within attraction basins, 
and joined together through their rapid quasi-radial multiclump  collapse, 
where the relative angular momentum of the collapsing/merging clumps is low.
A second phase follows, with 
lower mass assembly and star formation rates, but with higher angular momenta. 
This phase shows a variety of assembly patterns: i) only minor mergers without further significant
SF bursts but a SF tail, ii) major mergers with secondary SF bursts, as the galaxy HD-5004A shows in Figures~\ref{sfr-mass} and ~\ref{snapshot}, 
and/or iii) secular evolution of the galactic disk, as is the case for  G-1578411, which was shown in an earlier study to form its young 
bulge through secular processes after $z=1$, at least partially driven by a bar \citep{Brook:2012}.

The sizes, shapes, kinematics, stellar ages and metal contents of the stellar populations formed in these distinct phases, can be nicely distinguished in simulated bulges.
Indeed, we have found that the young component tends to occupy a smaller volume, to have disk-like morphology 
(note that G-1578411 is rather triaxial),
to be more rotationally supported, to have roughly solar metallicities and sub-solar $\alpha$-element enhancements.
The old population, by contrast, is more spheroid-like, has sub-solar metallicities and larger $\alpha$-element enhancements.
On the other hand, no clear trend with age shows up in the S\'{e}rsic indices, a result in agreement with observations.
The stellar metal content as well as the [Mg/Fe] ratios of these bulges have segregated  distributions and, in some cases,
show two clearly distinguishably peaks corresponding to the old and young populations.
These stellar populations are also clearly segregated by their loci in the [Mg/Fe] vs [Fe/H] plots, where the old population has only a mild slope, 
while the slope of the young population appears steeper, with the two populations meeting at the knee.
This kind of behavior is expected given the different nucleosynthetic origin of $\alpha$ and Fe-group elements.
                                                      
These trends have been shown to be robust against the different codes used, which have differences in their gravitational and hydrodynamical force integrators,
in  mass and spatial resolutions, simulation box sizes, star formation parameterizations, chemical feedback and evolution, and energetic feedback implementations. 
Also, by changing the bulge identification criteria from a simple radial cut to a kinematic based one, did not affect the age tendencies described above. 
The temporal cut-offs used to separate the two stellar populations have been chosen based on the shapes of the galaxy mass aggregation tracks and on the 
behavior of the SFRs at the bulge scale. Varying slightly these cut-offs does not affect these tendencies, either.   

If we associate the old populations formed during the rapid phase with classical bulges, and the young ones formed during the slow phase with pseudo bulges, 
all the simulations in this paper show both classical and pseudo bulge components, but with varying relative masses. 
Therefore, the classical vs pseudo-bulge characteristics would be a question of degree, rather than nature.
Otherwise, in all cases, measuring the light will result in higher relative contributions of the slow  phase of bulge formation, i.e., pseudo bulges. 
In this respect, we note that the feedback in G-1536 is particular effective at quenching star formation in the early phase, 
resulting in a relatively less significant (in mass) classical bulge population (see Table~\ref{tab2}). 
This may be pointing toward the solution to the issue raised in \citet{Kormendy:2010}, who point out that a significant number of local, 
massive spiral galaxies have pseudo bulges rather than classical bulges.

These assembly patterns are reminiscent of the two phases found in hydrodynamical simulations by \cite{DT:2006,Oser:2010,DT:2011} for more massive early type galaxies. 
The main difference lies in the percentage of gas transformed into stars at early epochs. 
In the case of massive ellipticals most of the available gas at the attraction basins for mass flows is transformed into stars during its "collapse" along the first phase.  
On the contrary, no such exhaustive gas consumption occurs for the less massive galaxies,
where gas remains available along the slow phase, first as a diffuse component, and later on being part of different structures 
(the host galaxy disk itself, other small disk-like satellites and/or small clumps) before it is incorporated to the bulge.

If indeed real bulges follow a similar formation pattern as simulated ones, their observational features discussed in Section \ref{intro} can be nicely explained. 
More so, this approach provides a possible explanation for some apparently paradoxical observational results.  
For example,  metal gradients could result from the different space distributions of the old and young populations, 
with the latter being more concentrated at the center due to dissipation. 
Also, bulge rejuvenation can be easily explained within this scenario.

We conclude that bulges can follow different assembly patterns, which can be summarized as two-phase processes 
(as in ellipticals) where non exhaustive gas transformation into stars occurs in the fast phase (unlike in ellipticals), 
with the additional effects, along the slow phase, of major and minor mergers, as well as of disk secular instabilities, in some cases.
These different patterns and their combinations in different epochs as well as in different proportions, might
explain the important dispersion in bulge properties observationally found.

The assembly of bulges is driven to a large extent by dynamical processes at larger scales. This is particularly true concerning bulge-forming starbursts.
As our simulations show, the bulge mass aggregation is a delayed consequence of its host galaxy mergers in the slow phase, and the result 
of collapse-like events in the fast one.
We have shown that galaxy mass and feedback can affect the relative contributions of these two phases, 
although it is certain that specific basin deformation/collapse processes, as well as merger histories will also play a role. 
A set of statistical samples of simulation runs with various physical parameters is needed in order to break the degeneracy 
with the merging history and provide further insights into the importance of the two phases of bulge formation.

\acknowledgments

We thank J. O\~norbe for sharing with us his analysis  software and  Patricia S\'{a}nchez-Bl\'{a}zquez for useful discussions.
We would also like to thank the anonymous referee for his constructive comments.  
This work was partially supported by the MICINN (Spain) through the grants
AYA2009-12792-C03-02 and AYA2009-12792-C03-03 from the PNAyA, as well as by 
the regional Madrid V PRICIT program through the ASTROMADRID network 
(CAM S2009/ESP-1496) and the ''Supercomputaci\'on y e-Ciencia'' 
Consolider-Ingenio CSD2007-0050 project.
We also thank the computer resources provided by BSC/RES (Spain)
and the Centro de Computaci\'on Cientif\'ica (UAM, Spain).
A. Obreja is supported by a FPI fellowship 
from MINECO, Spain, through the PNAyA.

\bibliography{aobreja}

\end{document}